\documentclass[10pt]{llncs}
\usepackage{amsmath}
\usepackage{indentfirst} 
\usepackage{caption}
\usepackage{subcaption}
\usepackage{graphicx}
\usepackage{xcolor}
\usepackage{multirow}
\usepackage{wrapfig}
\usepackage{epstopdf}
\usepackage{setspace}
\usepackage{algorithm}
\usepackage{algorithmicx}
\usepackage{algpseudocode}
\usepackage{color}

\algnotext{EndFor}
\algnotext{EndIf}
\algnotext{EndWhile}

\begin{document}

\title{Sibelia: A scalable and comprehensive synteny block generation
tool for closely related microbial genomes}
\author{Ilya Minkin, Anand  Patel, Mikhail
Kolmogorov, Nikolay Vyahhi, Son Pham}
\institute{Department of Computer Science and Engineering, UCSD, La Jolla, CA, USA.\\
St. Petersburg Academic University, St. Petersburg, Russia.}
\maketitle

\algnotext{EndFor}
\algnotext{EndIf}
\algnotext{EndWhile}
\begin{abstract}
Comparing strains within the same microbial species has proven
effective
in the identification of genes and genomic regions responsible for 
virulence, as well as in the diagnosis and treatment of infectious diseases. 
In this paper,
we present Sibelia, a tool for finding synteny blocks in multiple closely related
microbial genomes using \emph{iterative de Bruijn graphs}. Unlike most other tools,
Sibelia can find synteny blocks that are repeated within genomes as well as 
blocks shared by multiple genomes.   It represents synteny blocks
in a hierarchy structure with multiple
layers, each of which representing a different granularity level. 
Sibelia has been designed to work efficiently with a large
number of microbial genomes; it finds synteny blocks in 31 \textit{S. aureus} genomes within 31 minutes
and in 59 E.coli genomes within 107 minutes on a standard desktop.  Sibelia
software is distributed under the GNU GPL v2 license and is available at: 
https://github.com/bioinf/Sibelia. Sibelia's web-server is available at: http://etool.me/software/sibelia. 

\end{abstract} 
\vspace{-0.5cm}
\section{Introduction}

Early in the genomic era, sequencing a single representative isolate was thought to be 
sufficient to describe the genetics of a microbial species, and due to computational and technological 
limitations, \emph{comparative genomics} was restricted to comparing closely related microbial 
species. However, outbreaks of
virulent forms of common microbes (e.g. \textit{Escherichia coli O157:H7}) and multidrug-resistant 
bacterial strains (e.g. TB, MRSA) have  intensified efforts to understand
genetic diversity among microbial isolates belonging 
to the same species.

The task of decomposing genomes into non-overlapping highly conserved segments called
\emph{synteny blocks} has proven to be important in genome comparison. 
It has been applied to finding structural variations between genomes~\cite{kaper2004pathogenic,chambers2005community} and is also a prerequisite in most genome rearrangement 
software. Additionally, finding highly conserved regions shared
among many strains within the same microbial species 
helps to infer the minimum genomic material (or core genome)
required for  bacterial life and thus can be useful for the Minimal
Genome Project~\cite{gibson2008complete}.

Finding synteny blocks in multiple microbial genomes
 presents the following four challenges. (1) Genomes of strains 
belonging to the same species differ by point 
mutations, small/large indels, small-/large-scale rearrangements and duplications.
(2) With the current deluge of microbial genomes, 
genome comparison tools face the problem of comparing 
 hundreds or even thousands  of genomes simultaneously. A number of synteny block
generation tools exist~\cite{angiuoli2011mugsy,blanchette2004aligning,darling2004mauve,Pham2010drimm}, but 
most of them require the calculation of  
local alignments between all pairs of genomes. As
the number of genomes increases,
the number of required pairwise comparisons quickly becomes a bottleneck in terms of total
computational time.
(3) The task of synteny block reconstruction has been heavily dependent on parameters, which determine
the size and granularity (coarse-grained or 
fine-grained) of the resulting synteny
blocks~\cite{sinha2007cinteny}. Different
applications favor different scales of synteny block
reconstruction. For instance, while most current ancestral genome
reconstruction software~\cite{alekseyev2009breakpoint} 
favors large-scale
synteny blocks,
the analysis of virulence factors in pathogen genomes considers both 
small- (transposons, insert elements) and large-scale synteny blocks to be important~\cite{brussow2004phages}. Thus, general
synteny block
generation software should be able to find and represent synteny blocks
in multiple resolutions. (4) Synteny block software designed for a large number of bacterial genomes should be able to work
directly with a high volume of unannotated genome
sequences (represented in the alphabet of nucleotides) rather than
with annotated genomes (represented in the alphabet of genes).  This is because the accumulation of errors in gene annotation~\footnote{Errors in gene annotation can also be caused by using different software to annotate different genomes.} can grow substantially as the 
number of genomes increases. 

By concatenating multiple sequences into a highly repetitive ``virtual genome'', Peng et al.~\cite{peng2009decoding} noticed
that the problem of constructing
synteny blocks from multiple genomes is equivalent to the problem
of de novo repeat classification in the ``virtual genome'', and they utilized A-Bruijn graphs~\cite{pevzner2004novo} for synteny block reconstruction.

Since repeats are inexact, A-Bruijn graph frameworks require an initial step of
graph simplification to determine the consensus of repeats,
and later, \emph{threading} the genome through the simplified graph to determine the positions of repeats. The threading procedure
is usually problematic, and Peng et al.~\cite{peng2009decoding}
concluded that threading is a major bottleneck in synteny
block reconstruction.
Pham and Pevzner~\cite{Pham2010drimm} introduced the first A-Bruijn graph approach (DRIMM-Synteny) that does not
require the threading step by using a \emph{sequence modification algorithm}. Rather than simplifying the graph, this approach
modifies the sequence so that its corresponding graph is
simplified, and thus completely bypasses the threading procedure. As a result, 
the sequence modification procedure returns a sequence that is modified so
that its A-Bruijn graph reveals  synteny blocks as non-branching paths.

DRIMM-Synteny was designed to work with mammalian genomes, but it
faces the roadblock of constructing synteny blocks for large numbers of 
bacterial genomes because it takes as input a set
of genomes represented in the alphabet of genes. Adapting DRIMM-Synteny to work with raw DNA sequences faces many computational challenges described above.

Even when computational resources do not pose a problem, DRIMM-Synteny
and most other current synteny block generation tools 
do not present synteny blocks in order to
satisfy many different applications, namely, presenting synteny blocks in multiple granularity levels (i.e., different resolutions). 

Indeed, repeats in the ``virtual genome'', which are obtained by the concatenation of dozens to hundreds of simple
bacterial genomes, are both multi-scale (i.e., multiple size) and multi-granular, since
repeats in the virtual genome are accounted for not only by repeated blocks
within each bacterial genome, but also by blocks shared among multiple bacterial genomes.
Whereas the repeat size within each bacterial genome can range from dozens to several
thousands of bp, regions conserved among different genomes can have an even wider range of sizes; 
some of these regions may even reach several Mbp in size, and they usually contain sub-repeats. 
 For that matter, repeats are not exact, and
the longer the repeat, the more likely it is disrupted by
other smaller insertions/deletions. Thus, repeats of different sizes usually have different
granularities. 


While the de Bruijn graph has offered the best model
for representing perfect repeats in a simple genome~\cite{pevzner2004novo},
we argue that a single de Bruijn graph is not sufficient to capture the complicated repeat
structure of  virtual genomes (obtained by concatenating dozens to hundreds of simple genomes), which
are both multi-scale and multi-granular. In this work, we propose an iterative
de Bruijn graphs algorithm, which
uses multiple de Bruijn graphs constructed from different values of $k$ to capture the 
complicated repeat structure of  virtual genomes. 
Our iterative de Bruijn graphs algorithm
allows us to construct synteny blocks and represent them in a hierarchy structure. Large-scale (coarse-grained) synteny blocks
can be further decomposed into multiple layers, where each layer represents 
a different granularity level.

Our algorithm has been developed into 
Sibelia software, which offers a tool for decomposing multiple closely related
microbial genomes into synteny blocks. Sibelia has 
three special properties: (1) Sibelia is able to reveal synteny blocks repeated within 
genomes as well as blocks shared simultaneously by many genomes  (repeats within genomes are usually problematic for most current tools). (2) Sibelia represents
synteny blocks in a hierarchical structure. (3) Sibelia is fast:
it analyzes 31 \emph{S. aureus} genomes within
 31 minutes and 59 \emph{E.coli}
genomes within 107 minutes on a standard desktop. 

\vspace{-0.5cm}
\section{Methods}

For simplicity of 
exposition, we assume that the given set of genomic
sequences is concatenated (using delimiters) into a single 
``virtual'' genome and
consider the problem of finding syntenic (repeated) blocks 
within this \emph{highly duplicated genome}. From now on, we will
use the terms \emph{repeated block} and \emph{synteny block} interchangeably.

\subsection{De Bruijn Graph and Cycles}

Let a genome of length $n$ be represented as a circular
string $S=s_1\ldots  s_n$ over the nucleotide alphabet
$\{A, T, C, G\}$. A $k$-mer is a string of length $k$.
The de Bruijn graph $DB(S,k)$ represents every $k$-mer in $S$ as a vertex and connects
two vertices by a directed edge if they correspond to a pair 
of consecutive $k$-mers in the genome (these two $k$-mers overlap
in a shared $(k-1)$-mer).
The de Bruijn graph can be viewed as both a 
multigraph (i.e., adjacent vertices can be connected by multiple edges)
and a weighted graph with the multiplicity of an edge $(a,b)$ defined
as the number of times that the $k$-mers $a$ and $b$ appear consecutively in $S$.

\begin{figure}
\begin{center}
	\includegraphics[width=\linewidth]{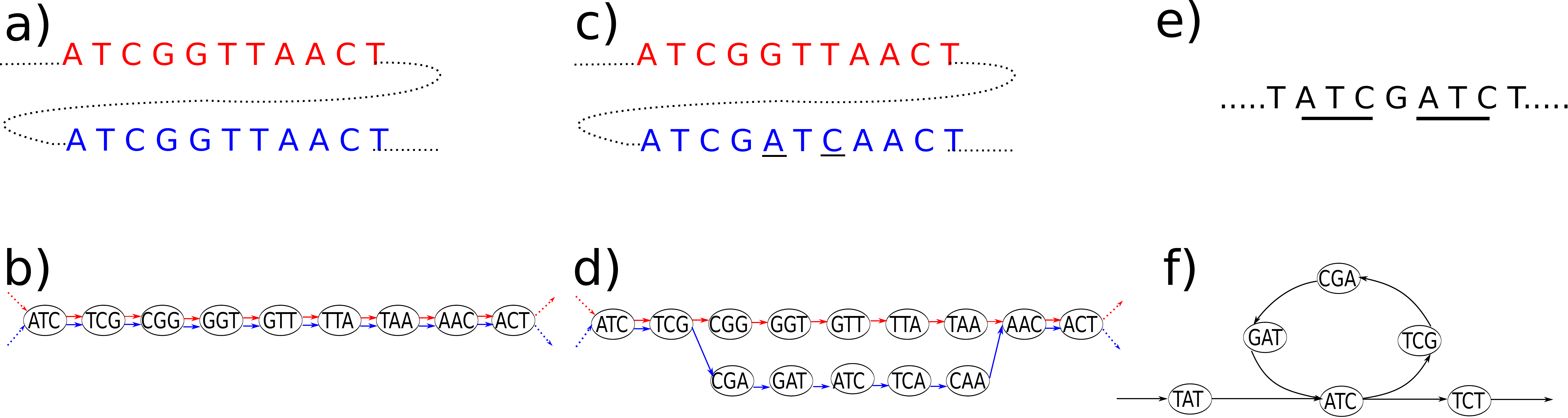}
	\caption{De Bruijn graphs and cycles in de Bruijn graphs. 
    a) A sequence $S = ATCGGTTAACG ... ATCGGTTAACG$ with 
the segment ATCGGTTAACG repeated twice. b) The de
Bruijn graph of the sequence in a). c) Sequence with inexact repeats: 
the green segment represents an inexact repeat of the red segment. 
Positions with different nucleotides are underlined. d) 
De Bruijn graph of the sequence in c). Minor differences between these two segments
(colored red and blue) are reflected by a bulge with two branches (the red and
the blue branches of the bulge). e) A sequence with a closely located
 repeated $k$-mer $ATC$. f) The de Bruijn graph of the sequence in e). The closely located
repeated $k$-mer ($ATC$) forms a loop in the graph.
}  
	\label{drawingNU}
\end{center}
\end{figure}

Alternatively, the de Bruijn graph of a string $S$ can be defined by a \emph{gluing operation} (see~\cite{Pham2010drimm,medvedev2011paired} for a
formal definition of this operation): represent $S$ as a
sequence of vertices $1,\ldots, n$ with $n-1$ edges $i \rightarrow (i+1), 1 \leq i \leq n-1$; label each vertex $i$ by the $k$-mer starting at position $i$ 
in $S$; \emph{glue} two vertices together if they have the same label (See Fig.~\ref{drawingNU} for de Bruijn graphs constructed from DNA strings).

Given a value of $k$ and a sequence $S$, perfectly repeated regions of size larger 
than $k$ in $S$ are \emph{glued} into paths in its de Bruijn graph $DB(S,k)$. Perfectly 
repeated regions that do not share any $k$-mer with other regions correspond to 
non-branching paths, which are maximal paths in the graph satisfying the condition that all their
internal vertices have only two neighboring vertices.  The multiplicity of a path is
equal to the number of times that the corresponding region appears in $S$.

One issue with using de Bruijn graphs for repeat analysis is that de 
Bruijn graphs constructed from real genomes have many short cycles and ``hide'' the 
genome's repeat structure. 
Cycles in de Bruijn graphs are commonly classified into two types: bulges (Fig.~\ref{drawingNU}d) and
loops (Fig.~\ref{drawingNU}f). Intuitively, bulges are caused by 
mismatches/indels between two
homologous sequences, and loops are caused by
closely located $k$-mer repeats. To reveal repeats in de Bruijn graphs, these small cycles
should be removed. To avoid the threading procedure, which is usually problematic, we adopt
a sequence modification approach to remove bulges in the de Bruijn graphs.
As will be made
obvious in the next subsection, Sibelia does not need to explicitly remove loops, but focuses only on bulges.  The reason behind this is that merely increasing the value of $k$ can help to eliminate
small loops in the de Bruijn graph. Fig.~\ref{drawingNU2}c shows a loop, which is 
caused by a closely located
3-mer repeat ($ATC$). The de Bruijn graph with larger vertex size $k =4$ does not
have a loop. Below, we formulate \textbf{SMP-B}: the sequence modification
problem for removing bulges in de Bruijn graphs.

We say that a string $P$ \textit{covers} an edge $a \rightarrow b$ in $DB(S,k)$ if $a$ and $b$ are consecutive
$k$-mers in $P$. A cycle $C$ in $DB(S,k)$ is classified as
a bulge if all its edges can be covered by two non-overlapping
substrings $P_1, P_2$ of $S$, and $P_1, P_2$ do not cover any edges in $DB(S,k)$ 
except for those in $C$.



\textbf{SMP-B}:\emph{ Given a string $S$ and parameters $\mathbf{C}$, $k$. Find a string $S'$
with minimum edit distance $d(S,S')$ such that $DB(S',k)$ has no two-way cycle
shorter than $\mathbf{C}$.}

Since the complexity of this problem remains unknown, we apply the \emph{sequence modification algorithm}~\cite{Pham2010drimm},
a heuristic algorithm for removing bulges. \\

\noindent
\textbf{Sequence modification algorithm}. Let $C$ be a bulge with total number of edges smaller than $\mathbf{C}$, formed by substrings $P_1$ and $P_2$
of $S$. 
To  remove the bulge, the
algorithm modifies $S$ by substituting 
all occurrences of $P_1$ in $S$ by $P_2$. 
Fig.~\ref{drawingNU2}a shows the de Brujn graph 
for $S = ATCG\mathbf{G}TTAACT...ATCG\mathbf{A}T\mathbf{C}AACT$, with two
inexact repeats. Minor differences between these two repeat instances
form a bulge having two branches (colored red and blue in the figure).
By changing $S$ into $S'=ATCG\textbf{G}TTAACT...A$ $TCG\textbf{G}TTAACT$ (i.e., substituting the blue
branch with the red branch), the bulge is also simplified. Note that $S'$ now 
contains an exact repeat of multiplicity 2 (Fig.~\ref{drawingNU2}b).

\subsection{Effects of $k$-Mer Size and Bulge Removal Procedure in Repeat Decomposition}
In this subsection, we give a relationship between repeats (revealed by non-branching paths) in de Bruijn graphs
constructed with different values of $k$ as well as repeats
revealed by non-branching paths before and after the bulge simplification procedure.  \\
\noindent
\textbf{Effects of $k$-mer size.}

\noindent
\textbf{Observation 1.}
Let $k_0, k_1$ be two positive integers such that $k_0 < k_1$, and let
$S$ be a cyclic genome. Furthermore, let $G_0$ and $G_1$ be the de Bruijn graphs
constructed from $S$ with $k= k_0$ and
$k=k_1$, respectively.  Any repeat $R$ revealed by a non-branching path in $G_1$ can be decomposed into a sequence of sub-repeats, each
corresponding to a non-branching path in $G_0$.

See Appendix for a formal description of this observation. Intuitively, when reducing the value of $k$-mer from
$k_1$ to $k_0$, some different non-branching
paths in $G_1$  having shared $k_0$-mers will
interfere with each other (these common $k_0$-mers play as new ``glues'' in the graph), thus, fragment large non-branching paths
into shorter ones. 
The above observation allows us to decompose any non-branching path of $G_1$
into a sequence of non-branching paths in $G_0$. In other words, 
repeats revealed by any non-branching
path from one de Bruijn graph can be
decomposed into a sequence of repeats (smaller sub-blocks) revealed by another de Bruijn graph constructed
from a smaller value of $k$.

\noindent
\textbf{Effects of Bulge Simplification.}

\begin{figure}
\begin{center}
	\includegraphics[width=1\linewidth]{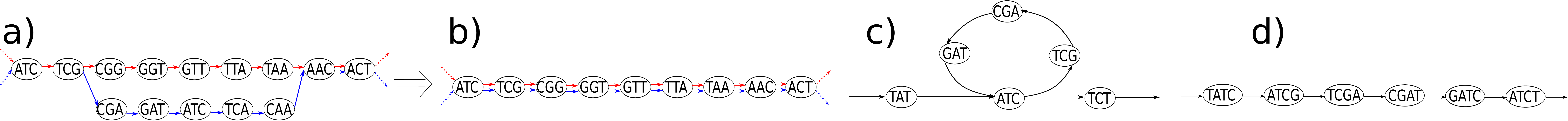}
	\caption{(a) De Bruijn graph of a sequence with two inexact repeats
        $S = ATCG\mathbf{G}T\mathbf{T}AACT ... ATCG\mathbf{A}T\mathbf{C}AACT$. The minor differences in the inexact repeats form
a bulge with two branches: The red branch $(TCG) \rightarrow (CGG) \rightarrow (GGT) \rightarrow (GTT) \rightarrow (TTA) \rightarrow (TAA) \rightarrow (AAC)$
  and the blue branch $(TCG) \rightarrow  (CGA) \rightarrow  (GAT) \rightarrow  (ATC) \rightarrow  (TCA) \rightarrow  (CAA) \rightarrow  (AAC) $. 
(b) We simplify the graph by changing the sequence from ATCGATCAACT to ATCGGTTAACT,
thus forming an exact repeat. The modified sequence corresponds to a
non-branching path on the de Bruijn graph. (c) A closely located repeated $k$-mer ($ATC$) forms a loop in the graph. (d) Increasing the value of $k$ can help to resolve the loop in c).}
	\label{drawingNU2}
	\vspace{-1cm}
\end{center}
\end{figure}

\noindent
\textbf{Observation 2.} 
Bulge simplification can be viewed as the process of ``merging'' consecutive
non-branching paths in the graph.  Fig.~\ref{drawingNU2}a shows a bulge, which
breaks the red segment into 3 segments, each corresponds to a non-branching
path. Fig.~\ref{drawingNU2}b shows the same graph after collapsing the bulge. Thus, the red large
repeat can be decomposed into 3 sub-segments; each of the subsegments corresponding to a 
non-branching path in the graph before simplification ($ATCGGTTAACT$ is decomposed into $(ATCG),\,(GTT),\,(AACT)$ according to Fig.~\ref{drawingNU2}).

These two observations are important for representing synteny blocks with different scales, which will be 
described in later sections.

\textbf{Challenges in Using de Bruijn Graphs in Repeats and Synteny Analysis}

Finding repeated blocks in a genome using de Bruijn graphs faces two 
problems. (1) Even in the case that synteny blocks are perfectly
repeated, they can have very different lengths. Representing 
both small (e.g., insertional elements, transposons) and large blocks in the de Bruijn graph is difficult because
if $k$ is set equal to a large value, then repeats of size smaller than $k$ can not be revealed, while
using a small value of $k$ may introduce additional gluings on large repeats. While
such additional gluings may help to reveal mosaic structures (subrepeats within a larger repeat)
of repeats, they also hide repeats when gluing is excessive. 
(2) Synteny blocks often contain many mismatches and
gaps, which restrict the use of large values of $k$ when constructing
the de Bruijn graph.

The first challenge motivates us to use different values of $k$ for representing repeats with different sizes. The second challenge
can be resolved by using the \emph{sequence modification algorithm}~\cite{Pham2010drimm} to remove bulges, since  sequence modification will eliminate 
mutations, gaps, and indels between homologous blocks and thus can help to increase the value of $k$.  We propose the \emph{iterative de Bruijn graph algorithm}
as follows.

Initially, the algorithm constructs the de Bruijn
graph from a relatively small value of $k= k_0$ and
performs graph simplification with a
small cycle length threshold ($C_0$). The
algorithm operates on the de Bruijn graph $G_0(S_0,k_0)$ and simplifies
all bulges using the sequence modification approach described above.
As a result, we
obtain a simplified de Bruijn graph $G_1$ and
the corresponding modified genome $S_1$. $S_1$ is
a distorted version of $S$ such that its de Bruijn graph $DB(S_1,k_0)$ does not
contain short bulges of length smaller than $C_0$. We note that graph simplification should be
applied using the \emph{sequence modification algorithm}; otherwise, $S_1$, the distorted 
version of $S_0$, is not available for the construction of the graph using a larger value of $k=k_1$.
The goal of the first iteration is
to collapse bulges caused by single point mutations or very short indels.
Thus, we can increase the value of $k$ and
construct a new de Bruijn graph $G_1 = DB(S_1,k_1)$, where
$k_1 > k_0$. The process continues until we reach a  value of $k$ that is
large enough to reveal large-scale synteny blocks (pseudo code of Sibelia is described in Algorithm 1) . Generally speaking, the iterative process should continue until the genome is presented as a single synteny block. 
This argument may appear unreasonable at  first sight, as our goal was to decompose genomes into synteny blocks. However,  one should
notice that the two observations in the above subsection allow us to retrace our steps to find previous synteny blocks.  

\subsection{Hierarchical Representation of Synteny Blocks}

\begin{figure*}[!t]
\center{\includegraphics[width=8cm]{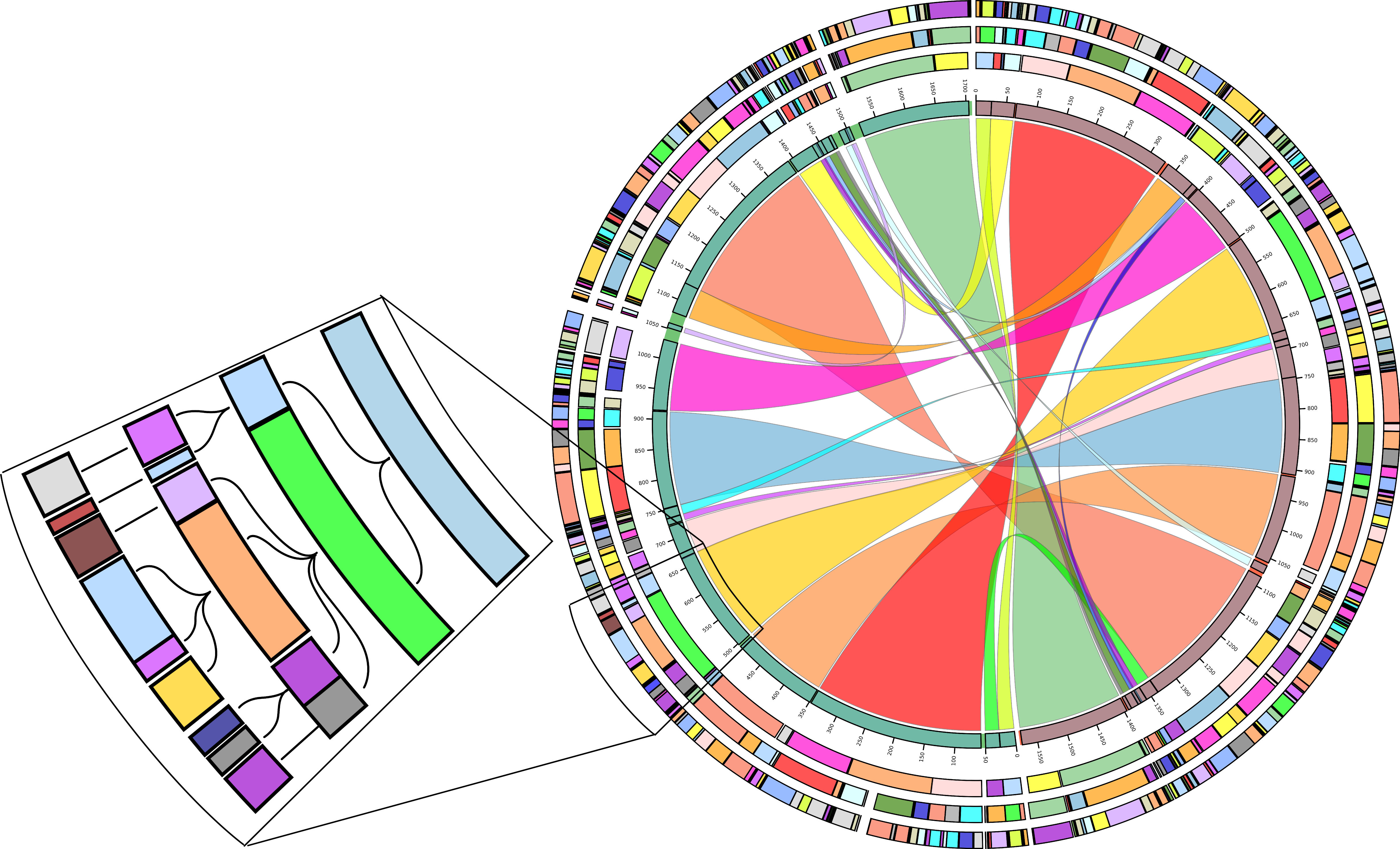}}
\caption{\small{The figure illustrates iterative construction of synteny blocks between two strains of Helicobacter pylori: F32 and Gambia94/24.
Each circle represents synteny blocks obtained at a particular stage. The outermost
circle represents the first stage of computation, the
next inner circle represents the second stage, and so on.
Synteny blocks are depicted by colored bands.
Multiple instances of the same synteny block
within each stage (circle) have the same color.
One can notice that from stage to stage,
blocks are merged together to form longer blocks. The panel on the left zooms in on a synteny block in the final stage. 
This panel depicts a tree that represents the decomposition of a synteny block into multiple layers.}}
\label{fig:hie}
\end{figure*}

Sibelia works iteratively as follows: from a sequence $S_{i-1}$,
it constructs a de Bruijn graph $G_i = DB(S_{i-1},k_i)$ and removes bulges to obtain a 
simplified graph $G_i^+ = DB(S_i,k_{i})$.  It then reconstructs the de Bruijn graph
for larger $k= k_{i+1}$: $G_{i+1} = DB(S_i,k_{i+1})$. Using Observation 1, each non-branching path in 
$G_{i+1}$ can be decomposed into a sequence of non-branching paths in 
$G_i^+$ because $k_{i+1} > k_i$. Since we only perform bulge simplification 
at each stage, each non-branching path in $G_i^+$ can in turn be
decomposed into a sequence of non-branching paths in $G_i$. Therefore, each 
non-branching path in the last stage can be represented as the root of a tree,
where its children are the decompositions of the previous stage. The leaves
of the tree represent  
non-branching paths in the de Bruijn graph
constructed from the smallest value $k_0$ (See Fig.~\ref{fig:hie} for the hierarchy representation of synteny blocks in two strains of \emph{H. pylori}).  
Under this decomposition, each chromosome can be considered as a single synteny block, which can be further decomposed into 
multiple large-scale ``synteny blocks''. Each large-scale synteny block can be further decomposed into smaller-scale synteny blocks. The process of decomposition continues until we reach the synteny blocks revealed by the graph $DB(S,k_0)$.

\textbf{Parameter choices}.
It appears that the iterative de Bruijn graph algorithm depends on 
many parameters: (1) the number of iterations; (2) for each iteration $i$, $k_i$ ($k$-mer size) and $C_i$ (cycle length for \emph{bulge simplification}).
However, we notice that the most important parameters
determining the outcome~\footnote{If the total coverage of synteny blocks for any set of genomes
within the same species is smaller than 40\%, then we classify the synteny block construction as unsuccessful according to the
analysis of core genomes in ~\cite{chattopadhyay2009high}.} of synteny block reconstruction 
are the values of parameters in the first iteration: $k_0$ and $C_0$. The
reason for this is that in microbial genomes, a point mutation event is the most common, and 
large indels occur at a much lower rate~\cite{lunter2008uncertainty}.    
In latter stages $I_i, (i > 0)$,  $k_i$ and $C_i$ reflect the size of repeated blocks and the granularity
of synteny blocks in that stage. While the choices of  $k_0$ and $C_0$ require a careful analysis of the 
evolutionary distance between genomes (See Appendix, Section 1), latter iterations can be seen as according users the
flexibility to choose granularity as well as the size of blocks in both the final and
intermediate stages (See Fig.~\ref{fig:hie} for a hierarchical representation of synteny blocks in two strains of \emph{H. pylori}).

Similar to the analysis in~\cite{chaisson2012mapping}, we derive the values for $(k_0,C_0)$ to be
$(30,150)$ for any set of microbial strains within the same species (see Appendix, Section 1 for a more detailed analysis). 
Sibelia's default mode has 4 stages (iterations) with the following parameters: $((30, 150), (100,1000), (1000,5000),$ and
$(5000,15000))$, where each pair of values corresponds to $(k_i,C_i)$ in the corresponding stage. While the pair of parameters for the first stage was derived using 
the sequence similarity of genomes within the same microbial species (see Appendix, Section 1), the final 3 stages ($k = 100, 1000, 5000$) were designed
to capture small repeats of length equal to several hundred bp ($k =100$),  transposons, insertion elements with average length about 1 Kbp~\cite{ohtsubo1996bacterial} ($k = 1000$), and 
large-scale blocks (usually comprising several genes) that are among multiple genomes ($k = 5000$). 
Users can add more iterations between any consecutive stages to obtain a ``smoother'' decomposition between stages.



%

\vspace{-0.5cm}
\begin{algorithm}
\caption{Iterative de Bruijn Graph}\label{euclid}
\begin{algorithmic}[1]
\Procedure{Iterative de Bruijn}{$G,((k_0,C_0), ..., (k_t,C_t) )$}
\State $S_0 \gets Concatenate(G)$ 
\State $Assert(k_0 < k_1 < ... < k_t)$
\State $Assert(C_0 < C_1 < ... < C_t)$

\State $i \gets 0$
\While{$i < t $}
\State $Graph_i \gets ConstructDeBruijnGraph(k_i,S_i)$ 
\State $S_i \gets SimplifyBulgesSmallerThanC(Graph_i,C_i)$
\State $i \gets i+1$
\EndWhile\label{euclidendwhile}
\State \textbf{return} $S_t, Graph_t$
\EndProcedure
\end{algorithmic}
\end{algorithm}

\vspace{-0.5cm}
\section{Results}
Since no gold standard exists for the evaluation of synteny blocks, we 
first benchmark Sibelia and other tools (Mugsy, Multiz, Mauve) on a simulated dataset.
The test case consists of two small hypothetical closely-related genomes, each 120 kbp long.
These genomes could be represented as permutations of synteny blocks as follows:
\begin{enumerate}
\item[] Genome \(1\): \(+4~~+2~~+3~~+1~~+3~~-4\) 
\item[] Genome \(2\): \(+5~~+2~~-3~~+1~~+3~~+5\)
\end{enumerate}

\begin{figure*}[!t]
\begin{center}
\subcaptionbox*{}{\vbox{\offinterlineskip\halign{#\hskip3pt&#\cr

\begin{subfigure}[b]{0.24\textwidth}
	\centering
	\includegraphics[width=\textwidth]{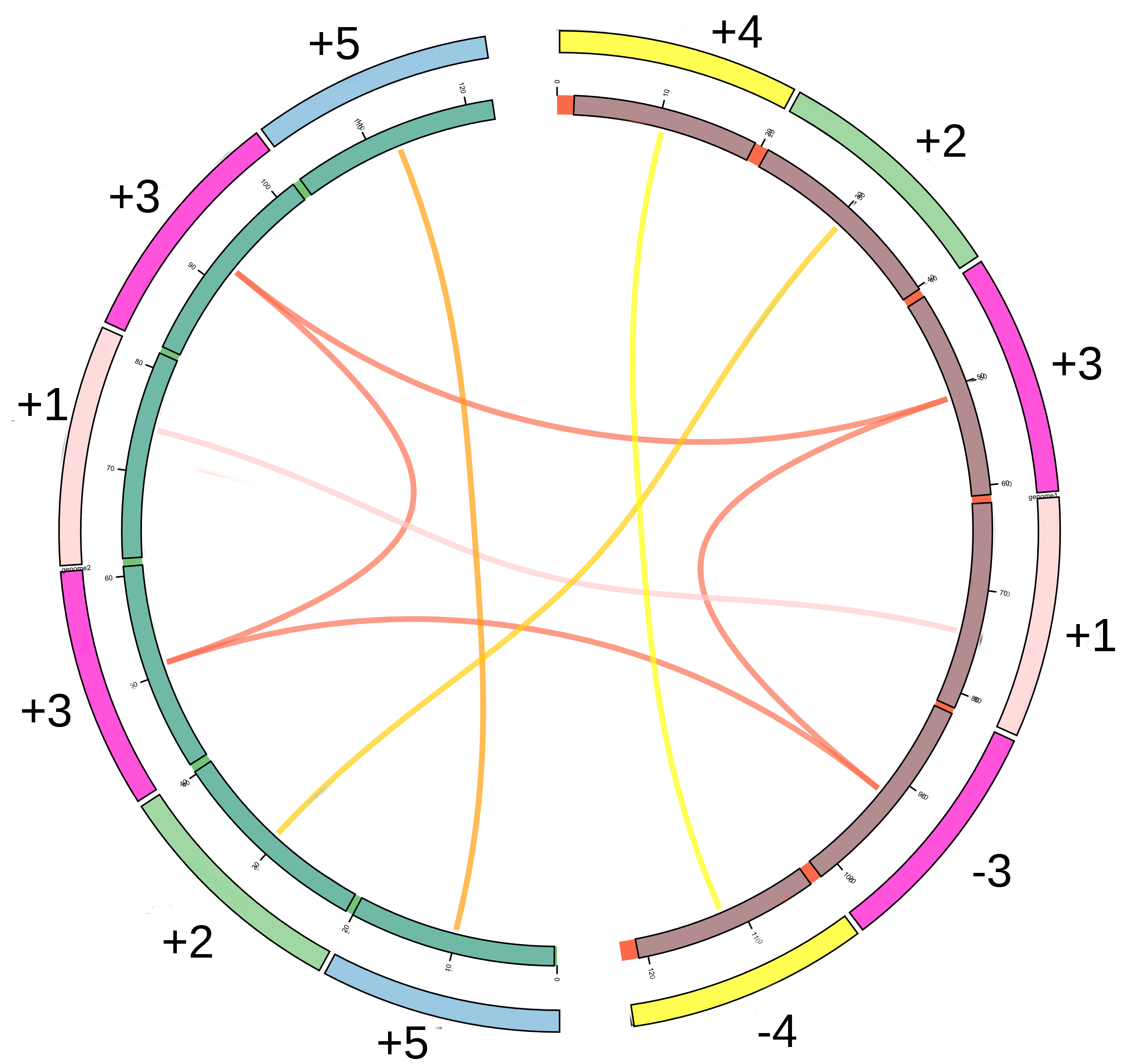}
	\caption{Sibelia}
\end{subfigure}
&
\begin{subfigure}[b]{0.24\textwidth}
	\centering
	\includegraphics[width=\textwidth]{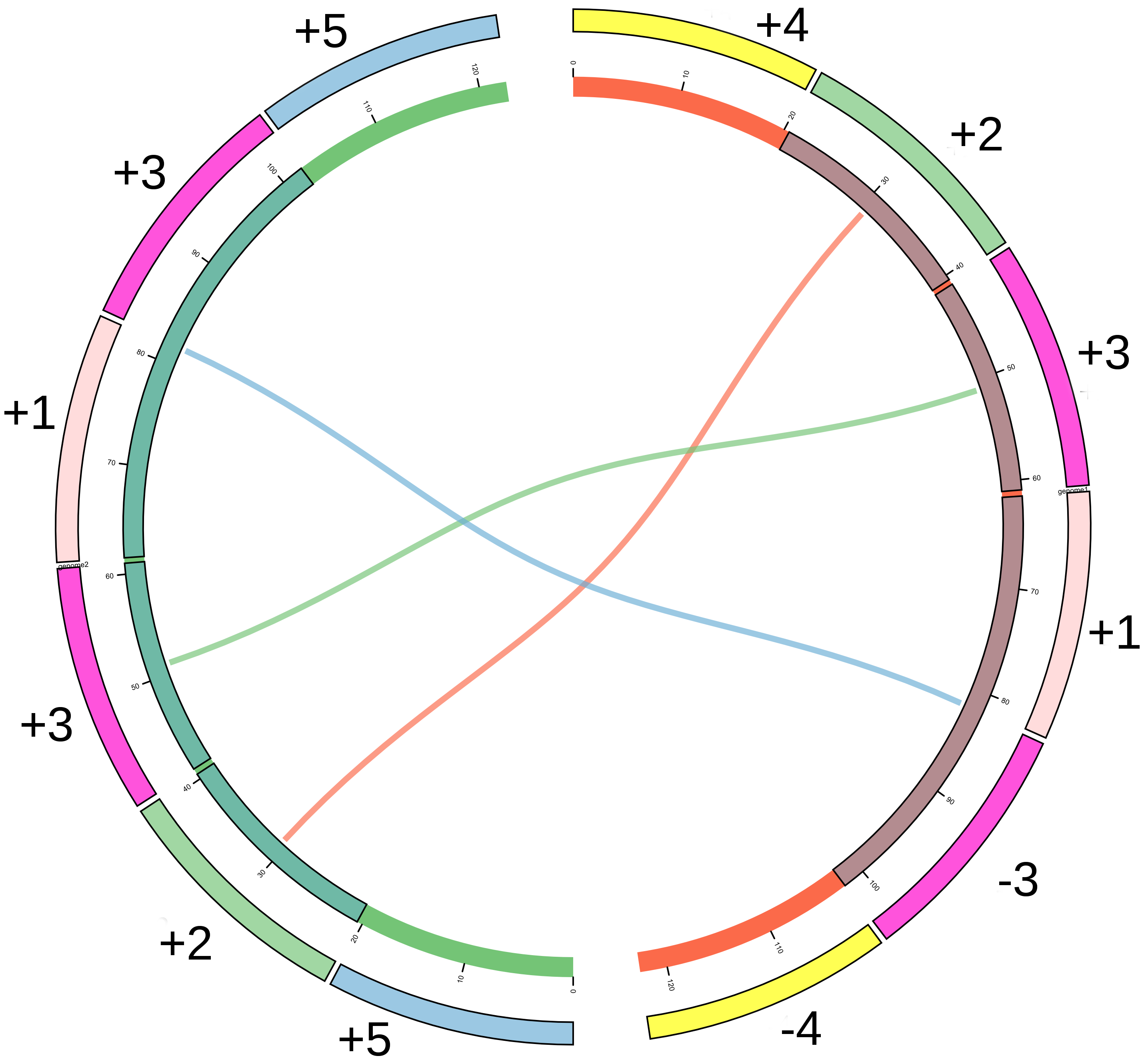}
	\caption{Mauve}
\end{subfigure}
\cr
\noalign{\vskip3pt}
\begin{subfigure}[b]{0.24\textwidth}
	\centering
	\includegraphics[width=\textwidth]{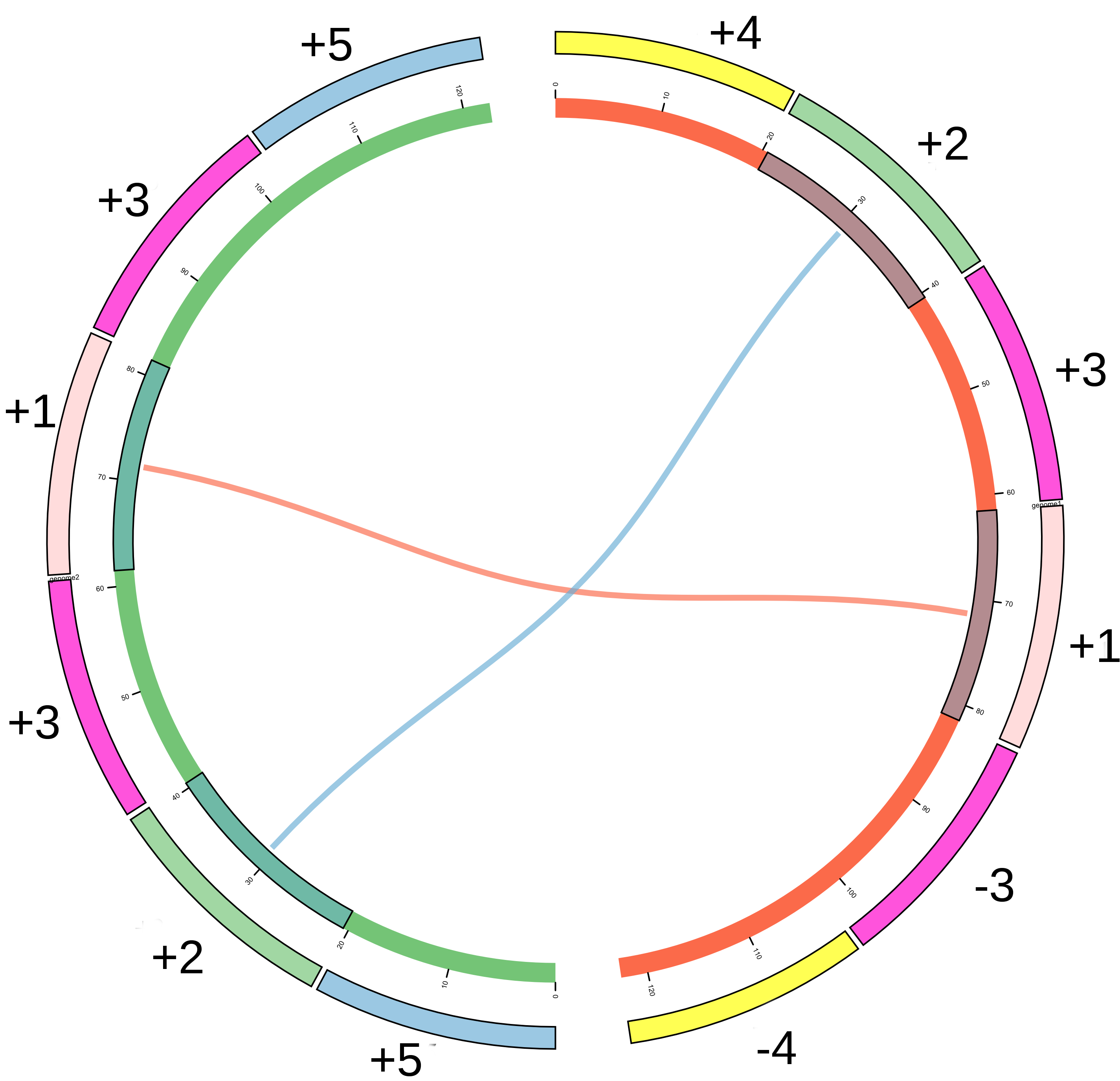}
	\caption{Mugsy}
\end{subfigure}
&
\begin{subfigure}[b]{0.24\textwidth}
	\centering
	\includegraphics[width=\textwidth]{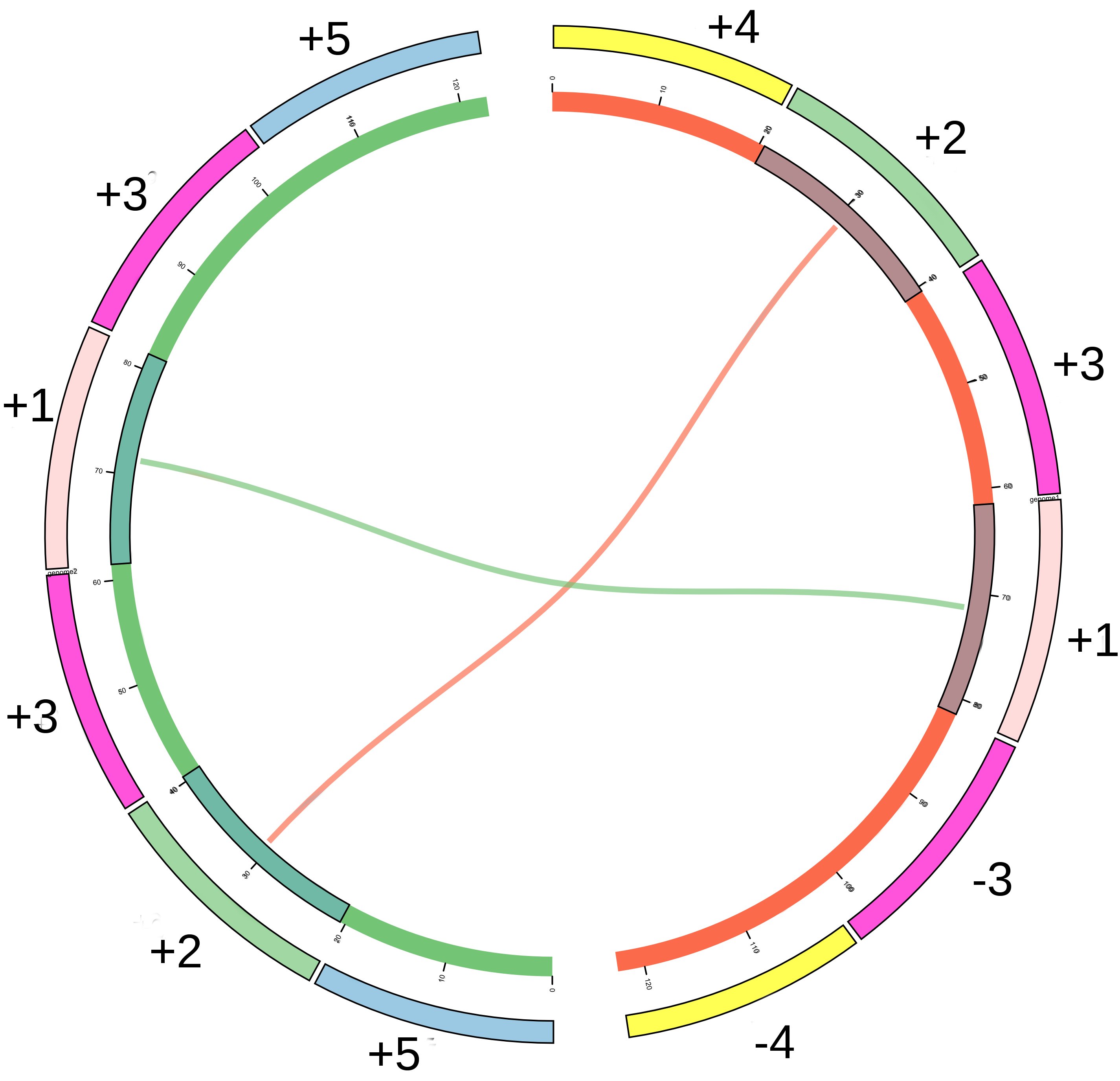}
	\caption{Multiz}
\end{subfigure}
\cr
}}}
\hfill
\subcaptionbox*{}{
\begin{subfigure}[b]{0.4\textwidth}
	\centering
	\includegraphics[width=\textwidth]{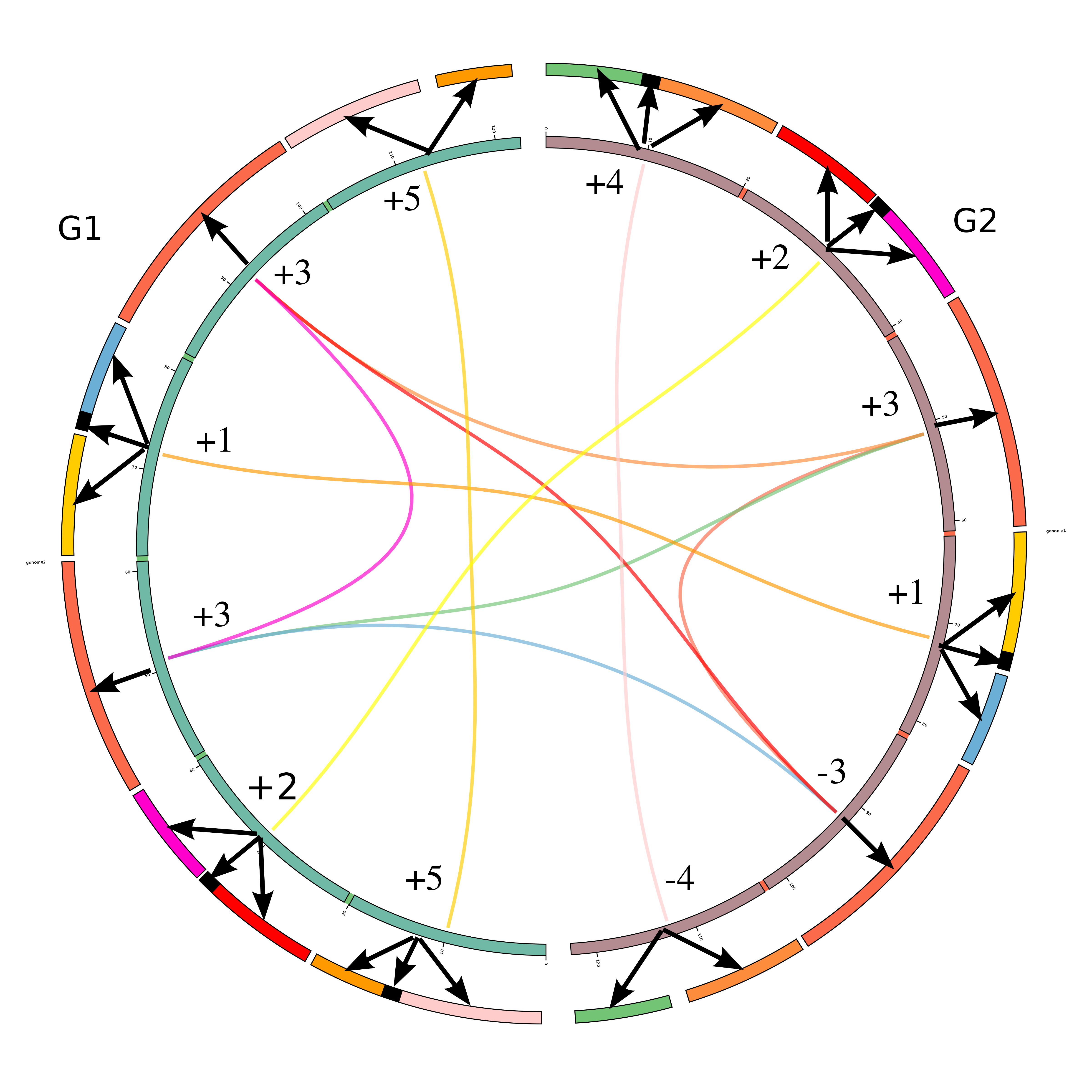}
	\caption{Sibelia -- A hierarchy representation of synteny blocks}
\end{subfigure}
}

\caption{Performance of different tools on  synthetic examples.
The outer circles in a), b), c), and d) indicate the gold-standard synteny blocks, where  
different instances of the same synteny block have the same color and are denoted by the same number.
Inner circles indicate blocks found by different tools.
All tools were run with their default parameters. a) The last stage from
Sibelia; b) Mauve ; c) Mugsy; d) Multiz. e) Sibelia represents synteny
blocks in a hierarchy structure. The inner circle indicates  
large-scale synteny blocks, and the outer circle shows synteny blocks at a finer scale (black blocks
represent insertion elements). Each large-scale synteny block (on the inner circle) corresponds
to the root of a tree, e.g., block +1 in 
$G_1$ (inner circle) can be decomposed into 3 blocks: yellow, black (insertion sequence), and blue
(outer circle).}
\label{offensivecase}
\end{center}
	
\end{figure*}

In the notation above, numbers depict synteny blocks, and signs designate their orientations.
All blocks are \(20\) kbp long.
These blocks indicate various types of repeats: blocks \(2\) and \(1\) correspond to common genetic cores,  
block \(3\) indicates a repeat common to both genomes, and blocks \(4\) and \(5\) are duplicated blocks within
each genome.
Different instances of a synteny block also contain point mutations, with a \(3\%\) probability for each position to change its nucleotide.

Fig. ~\ref{offensivecase} shows the results of different tools on the test case.
Sibelia correctly identifies all synteny blocks. No tool except Sibelia is able to locate blocks \(4\) and \(5\), and only Mauve detects repeats with multiplicity greater than 2 shared by both genomes.
Mugsy and Multiz rely on the Nucmer pairwise aligner package, thus limiting their ability to locate duplications within genomes.

We further demonstrate the ability of Sibelia in detecting and representing synteny blocks 
on multiple scales by making an additional complication to our simulated genomes. We generate a 
random DNA sequence of length 1,500 bp (a typical size of insertion sequences, which are common
in microbial genomes), and insert this sequence into some previous synteny blocks of these two genomes.
We also add 3\% mutations to each instance of the inserted sequence.

As different applications favor different synteny
block scales (e.g, MGRA~\cite{alekseyev2009breakpoint} may favor the
original decomposition, ignoring these insertion elements), other applications may find 
the translocation of these insert elements biologically significant and thus partition 
synteny blocks on a finer scale. Fig.~\ref{offensivecase}e shows the hierarchy presentation 
of Sibelia on the simulated example. In this figure, large-scale synteny blocks are
presented in the inner circle, while a finer representation of synteny blocks
(with insertion elements denoted as black blocks) is shown in the outer circle. Each synteny
block in the inner circle can be decomposed into a sequence of smaller synteny blocks in the outer circle.

\vspace{-0.4cm}
\subsection{Comparing Sibelia with Existing Tools}

We benchmarked Sibelia against Mugsy~\cite{angiuoli2011mugsy}, 
Multiz \cite{blanchette2004aligning},
and Mauve \cite{darling2004mauve} on 3 datasets: 
\emph{E.coli-3} --- 3 \emph{E.coli} genomes (15 MB), \emph{S.aureus-31} ---
31 \emph{S.aureus} genomes (90 MB), and \emph{E.coli-59} --- 59 \emph{E.coli} genomes (344 MB).
The first dataset \emph{E.coli-3} is used to demonstrate the quality 
of synteny block generation, while the other larger datasets show the 
memory consumption and running time performance of these different genome decomposition tools.  

\begin{figure}[h]
	\center{ \includegraphics[width=0.5\linewidth]{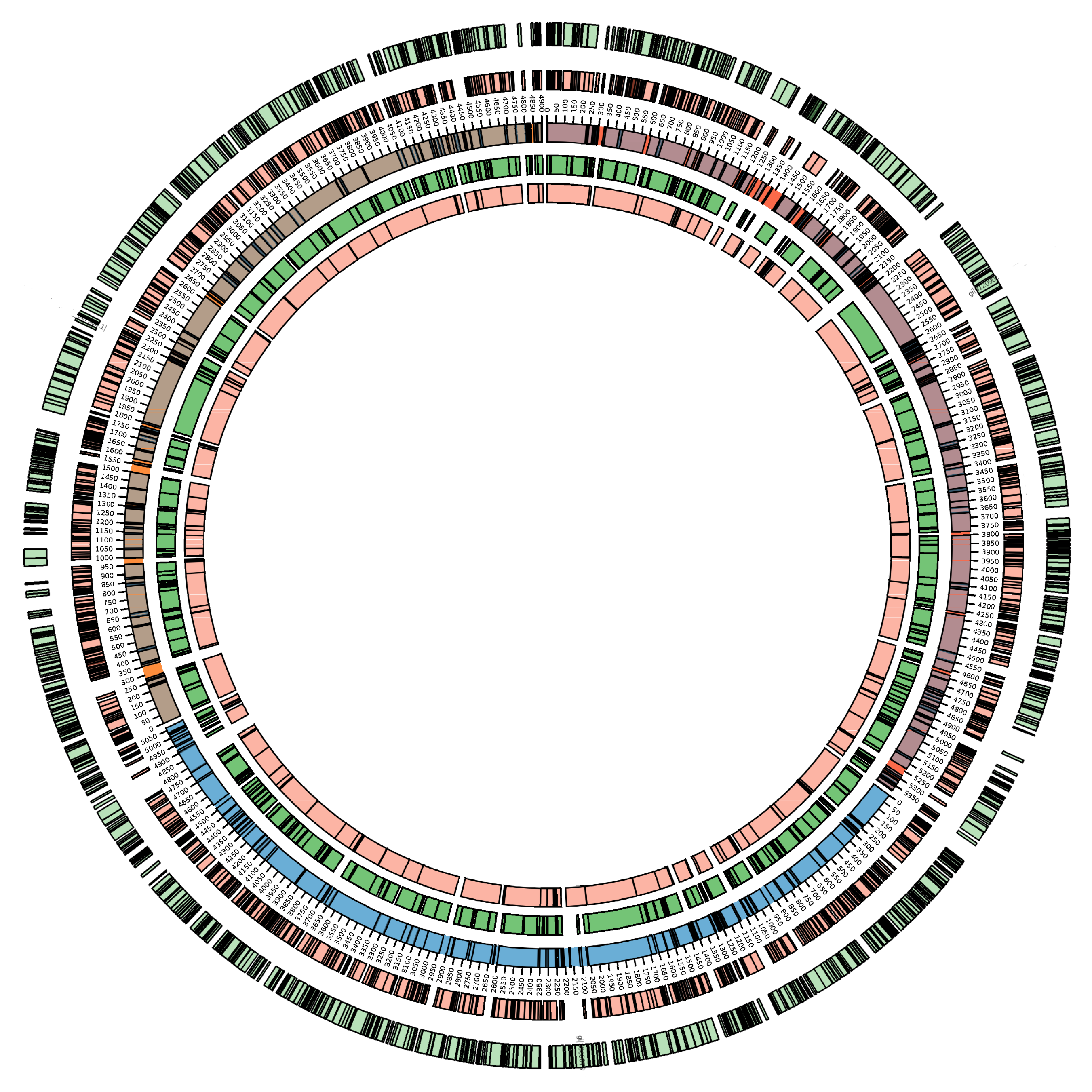} }
	\caption{\small{Circos diagram of synteny blocks on 3 \emph{E.coli} genomes. 
    From inside to outside: Mauve, Mugsy, Sibelia (the last stage), Multiz,
    and the first stage of Sibelia ($k=30, C= 150$). All tools were run with their
    default parameters.} }
	\label{fig:circos}
\end{figure}

On the \emph{E.coli-3} dataset, synteny blocks~\footnote{Mauve and Mugsy use the term ``locally
collinear block'' instead of ``repeated blocks".} generated from different tools are compared by 
\emph{genome coverage} and \emph{F-score}. We define the $F$-score of
synteny blocks generated from tools $T1$ 
and $T2$ as $F = 2(P R)/(P + R)$, where $P$ is
the fraction of nucleotides in the 
blocks reported by $T1$ that overlap with blocks reported by $T2$, and $R$ is the fraction 
of nucleotides in the blocks from $T2$ that overlap with blocks from $T1$
(see Table~\ref{tab:comparison}). The genome decompositions~\footnote{The ends of repeated blocks define
breakpoints on the genome and thus decompose the genome into segments of non-overlapping blocks}
of these tools are illustrated 
in Fig.~\ref{fig:circos}. While the genome decompositions from Sibelia, Mauve,
and Mugsy (shown by the three innermost circles in
Fig.~\ref{fig:circos}), are similar, Multiz's blocks are more fragmented. We do not 
criticize Multiz because different applications favor a different size and scale of repeated blocks.
Since Sibelia can present synteny blocks on multiple scales, we show  its genome decomposition
from the finest scale (first stage) in the outermost circle (Fig.~\ref{fig:circos}), which turns
out to be similar to Multiz's decomposition.

\begin{table}[!ht]	
\vspace{-0.5cm}
	\begin{center}
	\caption{Synteny block (LCB) Comparison}
	\label{tab:comparison}
	\begin{tabular}{ccccccc}
                    &  Genome Coverage & F-score \\ \hline
    Sibelia & 91 & 100 \\ 
    Mugsy & 82 & 95Â  \\ 
    Mauve & 90 & 95 \\ 
    Multiz& 70 & 85Â  \\
		\hline
	\end{tabular}
	\vspace{0.5cm}
        
        Sibelia, Mugsy, Mauve, Multiz were run with their default parameters.

	\end{center}
	\vspace{-0.5cm}
\end{table}

%
%
%
\begin{table}[!htb]
		\caption{Comparison of running-time/memory usage}
		\label{tab:runningtime}
		\begin{tabular}{ccccc}
			\hline
			& Sibelia & Mugsy & Multiz & Mauve\\
			\hline
			31 Aureus (min/GB) & 28/2.95 & 362/3.47 & 129/0.175 & 814/2.36\\
			59 E.coli (min/GB) & 107/8.75 & 749/9.23 & 815/0.6 & DNF/DNF \\
		\hline
		\end{tabular}
	\vspace{0.15cm}
	\begin{center}
	\caption*{The runtime and memory usage for all tools. All tools were run with default parameters. Tests were run on a single CPU
		Intel Xeon X5675 3GHz processor with 25GB RAM. 
		DNF: allocation error after 12 hours running.}
	\end{center}
\end{table}

As the number of compared genomes increases, Sibelia shows its advantage in running time performance.
When running on 59 \emph{E.coli} and 31 \emph{S.aureus} datasets, Sibelia proves to be 7 times
faster than Mugsy and Multiz on \emph{E.coli-59}, (see Table \ref{tab:runningtime}). The 
memory usage of Sibelia is similar to Mugsy and Mauve but is worse than
Multiz (Table~\ref{tab:runningtime}).  The synteny blocks that are shared 
among all genomes (59 \emph{E.coli} and 31 \emph{S.aureus}) cover $66.95\%$ and $54.25\%$ of the average of the genomes size. 
Using these synteny blocks, one can identify the core genome of each bacterial species. These numbers are consistent with 
the size of core genomes in \emph{S.aureus} and \emph{E.coli} previously reported~\cite{chattopadhyay2009high}. 

\vspace{-0.5cm}
\section{Discussion}
We have introduced Sibelia, a scalable and comprehensive new synteny block generation tool for analyzing
large numbers of microbial genomes belonging to the same species.  By using
the \emph{iterative de Bruijn graph}, Sibelia represents synteny blocks in 
a hierarchical structure that allows users to explore the composition of synteny blocks. We 
are aware that Cactus graphs~\cite{paten2011cactus} also decompose genome alignments into substructures based on
the topology of nested elements. Our algorithm of decomposing synteny blocks is different from the nested structure in the
Cactus graph, and we plan to further study the relation between these approaches. With the availability of Sibelia, 
studying genome rearrangement and genome evolution
using multiple levels of synteny blocks promises to be an interesting future research topic.


 
\vspace{-0.5cm}
\section{Acknowledgments} 
We would like to thank Pavel Pevzner, Hamilton Smith, Steve O'Brien, Alla Lapidus, Matt Schultz, Dinh Diep and Shay Zakov 
for many insightful discussions. We are indebted to Phillip Compeau, Nitin Udpa and Han Do for
revising the manuscript and for many helpful suggestions that
significantly improved the paper.  We would like to thank Hoa Pham for 
deploying Sibelia to the webserver.
This work was supported by the Government of the Russian Federation (grant 11.G34.31.0018)
and the National Institutes of Health (NIH grant 3P41RR024851-02S1).

\vspace{-0.5cm}
\section{Appendix}
\subsection{Microbial Species And The Choice of $k$}

While there is no uniquely accepted concept of species in bacteria,
the pragmatic species definition is based on DNA-DNA
hybridization (DDH)~\cite{wayne1987report}. According to this
definition, two isolates belong to the same species if they 
have $DDH > 70\%$, which
in turn corresponds to approximately $95\%$ average nucleotide
identity~\cite{konstantinidis2006bacterial}. In
other words, within a conserved
segment, each position has a 5$\%$ chance of
mutating.



These mutated points partition
any homologous region into 
a sequence of exact match segments with different lengths. Segments that 
are longer than $k$ (the size of a vertex in the de Bruijn graph)
are \emph{glued} together in the de Bruijn graph; we 
call these segments \emph{gluing
segments}. Two consecutive gluing segments correspond to a bulge
in the de Bruijn graph, and any non-gluing segments between the two consecutive gluing
segments correspond to branches of the bulge.
The distance between two consecutive
gluing segments characterizes the size of the bulge. 



The probability of encountering an exact matching region
of size $k$ is $P\{l=k\}= (1-\rho)^k\rho$, 
and the probability of encountering an exact matching region of size at least $k$ is
$P\{l\geq k\} = (1-\rho)^k$, where $l$ is the length of the exact matching region.
Given the value $\epsilon = 0.05$, 
the analysis in~\cite{chaisson2012mapping} allows us to characterize the function $d(k)$, which represents the distance
from a given position such that one can
encounter at least one gluing
segment (exact match segment with length
at least $k$) with probability $1-\epsilon$. According to~\cite{chaisson2012mapping}, 
$d(k) =  \frac{log(\epsilon)}{log(1-(1- \rho)^k)} (\frac{1}{\rho} - \frac{ k(1-\rho)^k}{1 - (1-\rho)^k})$. 
The function $d(k)$
characterizes the 
choice of bulge length threshold for simplification for each given value of $k$.

\begin{figure}
	\begin{center}
	\includegraphics[width=1\linewidth]{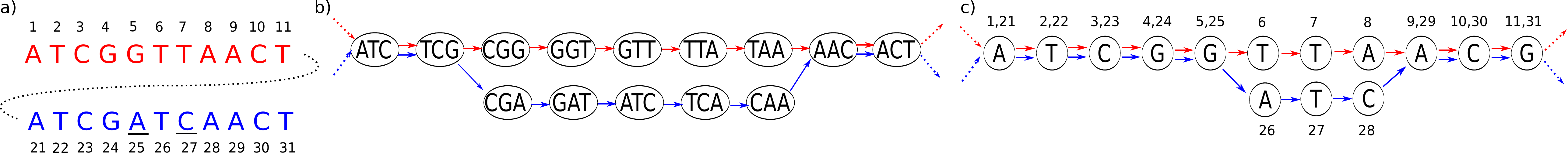}
	\caption{C-Graph. a) A genome sequence. b) De Bruijn graph for $k = 3$. c) C-Graph for $k=3$}
	\label{fig:cgraph}
	\end{center}
	\vspace{-0.5cm}
\end{figure}

\vspace{-0.5cm}
\section{Hierarchy Representation of Synteny Blocks}

\subsection{Parameter $k$ in repeats reconstruction}

In this subsection, we give a relationship of repeats that is revealed by non-branching paths in the de Bruijn graph
constructed with different values of $k$.  For the simplicity of proving the theorem, we introduce a different type of A-Bruijn graphs, 
called C-Graphs (Character Graphs), with a slighly different 
gluing rule from de Bruijn graphs.   
Given a value of $k$ and a string  $S$  of length $n$ formed over the alphabet $\{A,T,C,G\}$, the 
C-Graph $CG(S,k)$ is defined as follows:
\begin{itemize}
\item Represent $S$ as a graph with $n$ vertices labeled $1,\ldots n$ and $n-1$ edges $(i) \rightarrow (i+1)$. 
\item Glue vertex $i$ and $j$ if there exists $t \in [0,k]$ such that
$S[i-t:i-t+k-1] = S[j-t:j-t+k-1]$
\end{itemize}

Note that the de Bruijn graph can be obtained by changing the gluing rule 
above so that we glue $i$ and $j$ if $S[i:i+k-1] = S[j:j+k-1]$.

Each vertex $v$ corresponds to a set of integers $A(v)$, representing the positions that are glued
to this component. A position $i$ \emph{belongs} to a vertex if it is contained in $A(v)$. 
The C-Graph (See Fig.~\ref{fig:cgraph}) differs from the de Bruijn graph at the boundaries 
of repeats (branching vertices). The C-graph allows us to 
avoid overlapping synteny blocks at their shared branching vertices,
since each vertex is labeled by a single character that corresponds
to the character of $S$ at that particular position~\footnote{If multiple
positions are glued into the same vertex, we can also use the character to label any of these gluing positions, as they are identical}. 
The following theorem shows the relationship between synteny blocks revealed
by non-branching paths in C-graphs constructed from different values of $k$.

\begin{theorem}
Given two integers $k_0 < k_1$ and a cyclic genome $S$,
let $G_0$ and $G_1$ be the de Bruijn graphs
constructed from $S$ with $k= k_0$ and 
$k=k_1$, respectively. If $S[i:j]$ corresponds to a non-branching
path in $G_1$ (i.e., vertices $i$, $j$ belong to branching vertices and
there does not exist any $t \in (i,j)$ such that $t$ belongs to a branching vertex), 
then in $G_0$, $S[i:j]$ corresponds to a (not necessarily nonbranching) path
connecting two branching vertices containing $i$ and $j$.
\end{theorem}

\noindent \textbf{Proof}.

Since  $S$ corresponds to an edge-covering tour in a character graph, 
it's sufficient to prove that $i$ and $j$ belong to
branching vertices in 
$G_0$. Since $i$ belongs to a branching vertex in $G_1$, let
$I = \{i_1\ldots, i_r\} (i\in I)$ be a set of positions in $S$ that 
are glued to this vertex.  It is evident that these positions will also 
be glued together in $G_0$ because $k_0 < k_1$. Since $i$ belongs to a branching
vertex in $G_1$, there must exist $i_{t1}, i_{t2} \in I$  such that
either $S[i_{t1} -1] != S[i_{t2} -1]]$ or $S[i_{t1}+1] \neq S[i_{t2} +1]$. This
also implies that $i$ belongs to a branching vertex in $G_0$.
Similarly, we can prove that $j$ belongs to a branching vertex in $G_0$.



\begin{spacing}{0.95}
\bibliographystyle{splncs03}
\bibliography{mybibnew}
\end{spacing}
\end{document}